\documentclass[12pt]{iopart}
\usepackage{graphicx}
\usepackage{cite}
\usepackage[utf8]{inputenc}
\usepackage[T1]{fontenc}
\usepackage{url}
\usepackage{hyperref}
\usepackage{bm}
\usepackage{tabularx}
\usepackage{xcolor}
\begin{document}

\title[]{Hall effect on the magnetic reconnections during the evolution of a three-dimensional magnetic flux rope }

\author{K. Bora$^1$\footnote{Presently at Max Planck Institute for Solar System Research, Goettingen, Germany}, Satyam Agarwal$^{1,2}$, Sanjay Kumar$^3$, R. Bhattacharyya$^1$}

\bigskip

\address{$^1$Udaipur Solar Observatory, Physical Research Laboratory, Dewali, Badi Road, Udaipur-313001, India.}
\address{$^2$ Discipline of Physics, Indian Institute of Technology, Gandhinagar 382355, India.}
\address{$^3$ Department of Physics, Patna University, Patna-80005, India.}

\ead{kamleshb@prl.res.in}
\vspace{10pt}
\begin{indented}
\item[]September, 2022
\end{indented}

\begin{abstract}
We present a novel Hall magnetohydrodynamics (HMHD) numerical simulation of a three-dimensional (3D) magnetic flux rope (MFR)---generated by magnetic reconnections from an initial 3D bipolar sheared field. Magnetic reconnections during the HMHD evolution are compared with the MHD. In both simulations, the MFRs generate as a consequence of the magnetic reconnection at null points which has not been realized in contemporary simulations. Interestingly, the evolution is faster and more intricate in the HMHD simulation. Repetitive development of the  twisted magnetic field lines (MFL) in the vicinity of 3D nulls (reconnection site) is unique to the HMHD evolution of the MFR. The dynamical evolution of magnetic field lines around the reconnection site being affected by the Hall forcing, correspondingly affects the large-scale structures.
\end{abstract}

\maketitle

\section{Introduction}
Hall magnetohydrodynamics (HMHD) recognizes the importance of the Hall field $\textbf{J}\times\textbf{B}$ in the generalized Ohm's law where \textbf{J} is the current density and \textbf{B} is the magnetic field. At the ion inertial scale ($\delta_i$), the Hall term decouples the electron and ion motions, making magnetic field lines (MFL) to get frozen into the electron fluid. Consequently, field lines can slip through the ion fluid that accounts for the momentum change (electron mass being negligible). Such slippage of  field lines can change the field line connectivity between two ion-fluid parcels, amounting to magnetic reconnection \cite{Axford1985}. Moreover, the Hall field neither affects the magnetic energy nor the magnetic helicity dissipation rates and hence, preserves the magnetic topology \cite{Hornig}. HMHD is believed to expedite the reconnection rate and accounts for the impulsive nature of the explosive solar coronal transients such as solar flares, coronal mass ejections (CMEs), and jets \cite{BhattacharjeeReview}. Hall effect also plays an important role in space plasmas like Earth’s magnetosphere---the magnetopause and the magnetotail reconnection where current sheets exist \cite{Mozer2002}. Moreover, a straightforward order analysis of the dimensionless induction equation presented in ref. \cite{Bora2022} emphasizes the inevitability of the Hall effects in solar coronal reconnections.

HMHD is well known to make any two-dimensional (2D) evolution 3D by the generation of an out-of-reconnection plane magnetic field component \cite{Huba, Bora2021}. In particular, starting from an initial 2D field, ref. \cite{Bora2021} has documented the formation of a 3D magnetic flux rope (MFR) due to the Hall magnetic field. Further, the same paper also analyzes Hall-assisted magnetic reconnection for an MFR activation initiated with a 2D bipolar sheared arcade. The study confirms the fast and impulsive nature of magnetic reconnection in HMHD and reveals intricate details such as local breakage of MFR owing to internal intermittent reconnections within the rope itself. Nevertheless, further research in this direction is required because the MFLs responsible for solar flares and CMEs are 3D owing to their inherent twist. 
Relevantly, in addition to being underlying significant structures to the solar eruptions, the MFRs also play a vital role in the dynamical evolution of the some
laboratory plasma systems where the involved length scales are around or below the ion inertial scale \cite{moser2012, van2012, shi2022}. It is then imperative to explore the HMHD evolution of an initially 3D magnetic configuration toward the creation and dynamics of MFRs. For the purpose, here we present a comprehensive analysis of 3D Hall-assisted magnetic reconnection during the evolution of an MFR.

The simulation results presented here complement the results obtained in ref. \cite{Bora2021}, where the formation of MFR through magnetic reconnections is demonstrated in an initial 2.5D magnetic field (having translation symmetry along $y$-direction). However, in contrast to ref. \cite{Bora2021}, here the initial magnetic field is constructed to support 3D bipolar sheared arcades without any translation symmetry---making the initial magnetic topology more realistic. The simulated dynamics show the development of magnetic nulls and subsequent reconnections at these nulls, which ultimately lead to the formation of an MFR.  Importantly, the MFR is found to be anchored at the bottom boundary. Such a process of reconnection via null formation  
is not realized in ref\cite{Bora2021}. Moreover, the anchored MFRs are observationally more relevant to solar coronal transients than the non-anchored levitating ones, as shown in refs. \cite{Sanjay2016, Bora2021}. 
Hence, the novelty of this work is in its demonstration of the MFR in the near-neighborhood of the newly formed null points---resulting from the local enhancement in the gradient of the magnetic field---leading to efficiently faster reconnections and more complex dynamics.

The rest of the paper is arranged as follows. The numerical model and initial setup are described in Section
2. The results of the HMHD and the MHD simulations are discussed in Section 3.
Section 4 summarises these results
and discusses the key findings of the paper.

\section{Numerical model and setup}
In order to ascertain the MFR generation, the simulations are initiated with a 3D bipolar sheared field $\textbf{B}^{*}$ \cite{Sanjay2016}. The $\textbf{B}^{*}=\textbf{B}+a_0 \textbf{B}^{'}$, where the components of $\textbf{B}$ are 
\begin{eqnarray}
B_x &= a_{z} \sin (a_{x}x) \exp\left(\frac{-a_{z}z}{s_{0}}\right)~,\\
B_y &= a_{y} \sin{(a_{x}x)} \exp\left(\frac{-a_{z}z}{s_{0}}\right)~,\\
B_z &= s_{0} a_{x} \cos{(a_{x}x)} \exp\left(\frac{-a_{z}z}{s_{0}}\right)~,
\end{eqnarray}

and the components of the $\textbf{B}^{'}$ are
\begin{eqnarray}
{B_{x}}^{'}  &= (\sin x \cos y - \cos x \sin y) \exp\left(\frac{-z}{s_0}\right)~,\\
{B_{y}}^{'}  &= -(\cos x \sin y + \sin x \cos y) \exp\left(\frac{-z}{s_0}\right)~,\\
{B_{z}}^{'}  &= 2 s_0 \sin x \sin y \exp\left(\frac{-z}{s_0}\right)~.
\end{eqnarray}

\begin{figure}[h]
\includegraphics[width=1\linewidth]{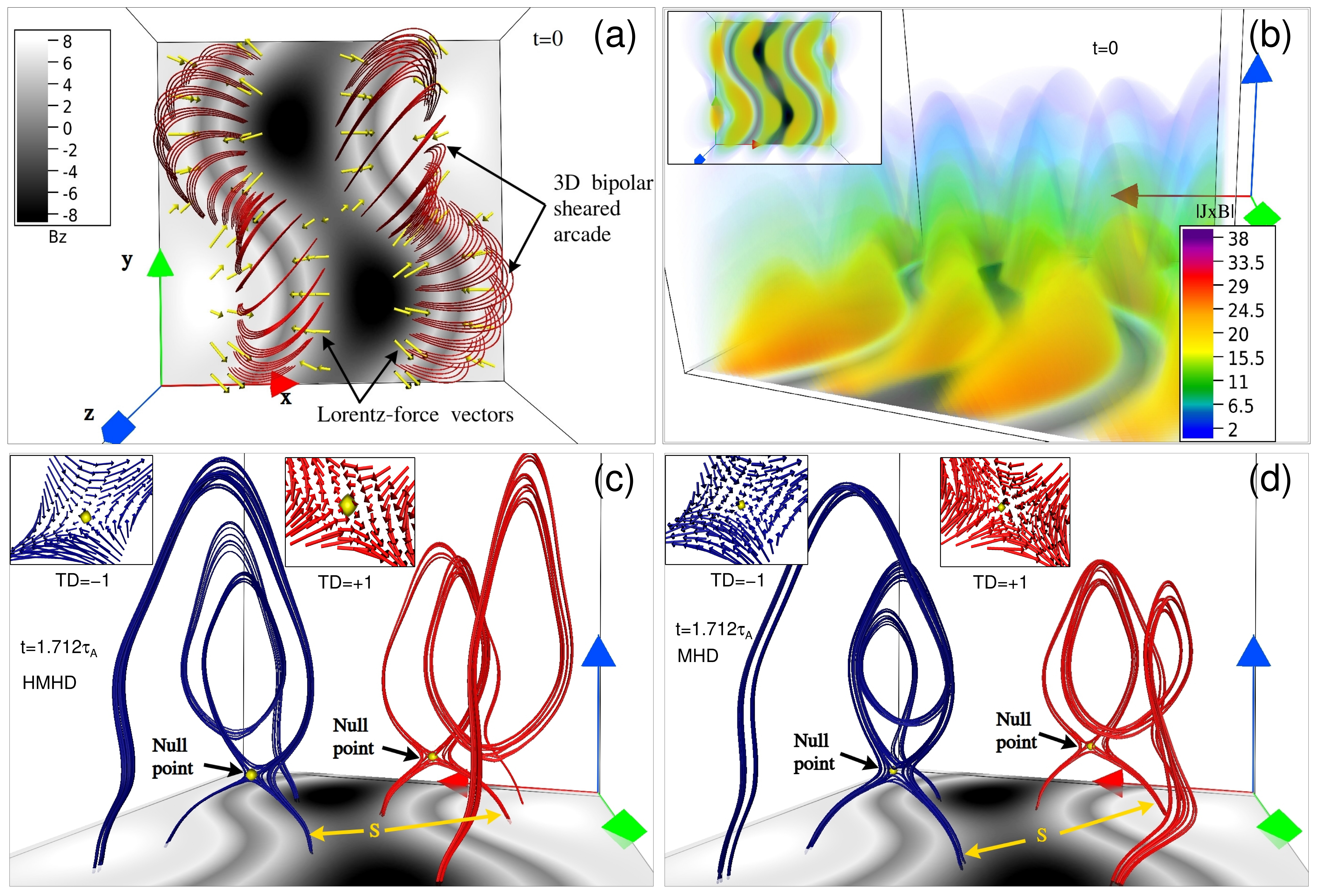}
\caption{\label{fig1} (a) Initial 3D sheared magnetic field lines (red) along with the oppositely directed Lorentz force vectors (yellow) around the sigmoid-shaped polarity inversion line ( PIL). (b) The magnitude of Lorentz force ($|\textbf{J}\times\textbf{B}|$) and its top-down view (inset image in the top left corner). (c) and (d) Two 3D nulls (yellow) of topological degrees -1 (blue) and +1 (red) during the HMHD and MHD simulations respectively. The spine is indicated by the yellow arrows. 
The bottom boundary in all the panels of this figure and subsequent figures shows the $B_z$ maps in grayscale, where the lighter shade represents positive-polarity regions and the darker shade indicates the negative-polarity regions. The red, green, and blue arrows in each panel represent the $x-$, $y-$, and $z-$ axis of the Cartesian coordinate respectively.}
\end{figure}

\noindent The parameters used in the simulation are given in Table \ref{tab1}. The $\textbf{B}^{*}$ has a sigmoid-shaped polarity inversion line (PIL) (see Figure \ref{fig1}(a); at PIL $B_z =0$). The MFLs of the $\textbf{B}^{*}$ are shown in Figure \ref{fig1}(a). The expression of the Lorentz force associated with the initial magnetic field is provided in Appendix A. The force falls
off sharply with height, such that the upper half of the computational domain is nearly in a force-free state (see Figure \ref{fig1}(b)). The initial Lorentz force is crucial in our simulations as it triggers the simulated dynamics. Dynamical evolution of the initial field is governed by the following set of equations \cite{Bora2022} 
 \begin{eqnarray}
\label{momtransf}
\frac{\partial{\bf v}}{\partial t} +({\bf v}\cdot \nabla){\bf v}&=&
 -\nabla p + (\nabla\times{\bf B})\times{\bf B} + 
\frac{1}{R_F^A}\nabla^2 {\bf v}~,\\
\label{induc}
\frac{\partial{\bf B}}{\partial t}&=& \nabla\times(\textbf{v}\times{\bf B})
-d_H\nabla\times((\nabla\times{\bf B})\times{\bf B})~,\\
\label{incompv}
\nabla\cdot {\bf v}&=& 0~, \\
\label{incompb}
\nabla\cdot {\bf B}&=& 0~,
\end{eqnarray}
where $R_F^A=(v_A L_0/\nu)$, $\nu$ being the kinematic viscosity, $v_A$ being the Alfv\`en speed (defined shortly), and $L_0$ being the characteristic length scale---is the fluid Reynolds number.
 The dimensionless equations (\ref{momtransf})-(\ref{incompb}) use the following normalization,
\begin{equation}
 \label{norm}
       {\bf{B}}\longrightarrow \frac{{\bf{B}}}{B_0},
  \quad{\bf{x}}\longrightarrow \frac{\bf{x}}{L_0},
  \quad{\bf{v}}\longrightarrow \frac{\bf{v}}{v_A},
  \quad t \longrightarrow \frac{t}{\tau_A},
  \quad p \longrightarrow \frac{p}{\rho_0 {v_{A}}^2}~. 
 \end{equation}
Here we assume arbitrary $B_0$ and $L_0$ while $v_A \equiv
B_0/\sqrt{\mu_0\rho_0}$ being the Alfv\'en speed and $\tau_A\equiv L_0/v_A$ being the Alfv\`en timescale. $\mu_0$ is the permeability of the vacuum, $\rho_0$ is a constant mass density, and $d_H=(\delta_i/L_0)$ is the Hall parameter where $\delta_i$ is the ion-inertial scale length. Equations (\ref{momtransf})-(\ref{incompb}) reduce to the MHD equations \cite{avijeet2018} when $d_H=0$.  Relevantly, the pressure $p$ (initially set to zero) satisfies the elliptic boundary value problem, generated by imposing the incompressibility constraint (\ref{incompv}) on the momentum transport Equation (\ref{momtransf}); see ref. {\cite{Bhattacharyya2010}} and the references therein. The plasma is assumed to be thermodynamically inactive \cite{Bora2022}.

\begin{table}[!h]
\caption{\label{tab1}List of parameters used in the simulations is given in the table below.  }
\begin{center}
\begin{tabular}{ccccc}
$a_x$& $a_z$& \mbox{$a_y$}& \mbox{$s_0$}& \mbox{$a_0$}\\
\hline
1.0& 0.9& \mbox{$\sqrt{{a_x}^2-{a_z}^2}$}& \mbox{6}& \mbox{0.5}\\
\end{tabular}
\end{center}
\end{table}

Equations (\ref{momtransf})-(\ref{incompb})
are numerically solved using the EULAG-MHD---a magnetohydrodynamic extension \cite{PiotrJCP} of the well-established Eulerian/Lagrangian comprehensive geophysical fluid solver EULAG \cite{Prusa}. 
EULAG is based on the MPDATA ({\it multidimensional positive definite advection transport algorithm})  which is a spatiotemporally second-order-accurate nonoscillatory forward-in-time advection transport algorithm \cite{Piotrsingle}. Relevant here is its
proven dissipative property, intermittent and adaptive to the
generation of under-resolved scales in field variables for a fixed
grid resolution. An unbound increase in field gradient is then smoothed out
by this MPDATA produced locally effective residual dissipation of the 
second order which is sufficient to sustain the monotonic nature of the solution. 
Consequently, magnetic field lines reconnect at locations of maximal current gradients 
having the length scale  $\mathcal{O} (\Delta \textbf{x})$---where $\Delta \textbf{x}$ is the spatial step-size. Such mimicking of the action of explicit subgrid-scale turbulence models wherever the concerned advective field is under-resolved is characteristic of the implicit large-eddy simulations (ILES) \cite{Grinstein2007}. Based on the earlier successful simulations  ref. \cite{SKRB, Sanjay2016, SK2017, Ss2020, Sanjay2021} here also we rely on the ILES functionality 
of MPDATA to onset magnetic reconnection. The residual dissipation being implicit, its characterization is 
not straightforward. Its quantification is meaningful only  in the spectral space where analogous to the eddy viscosity of explicit subgrid-scale models for turbulent flows, it only acts on
the shortest modes admissible on the grid \cite{Domaradzki}; in particular, in the vicinity of steep gradients in simulated fields. We keep this as a future task.

In this paper, we perform the HMHD and the MHD simulations initiated with the magnetic field $\textbf{B}^{*}$. The simulations are conducted by assuming the plasma to be incompressible, thermodynamically inactive, and explicitly nonresistive. A physical domain of extent [\{0,2$\pi$\}, \{0,2$\pi$\}, \{0,8$\pi$\}] is resolved by a computational domain of size 64$\times$64$\times$128, making the spatial
step sizes $\Delta x=\Delta y=$0.0997, and $\Delta z=$0.1979 (in dimensionless units). The simulations start with a motionless state, i.e. initial flow velocity field ($\textbf{v}$) is set to zero. The mass density $\rho_0$ is set to 1 and the Reynolds number $R_F^A$ is set to 50,000. 
All the parameters are the same for the HMHD and MHD simulations except $d_{H}$.
 For the MHD simulation the value of $d_{H}=(\delta_i/L_0)$ is set to 0. In conformity with the ILES,  $d_{H}$ is set to 0.04 for the HMHD simulation. The corresponding ion-inertial scale lengths are 
 $\delta_i\approx 0.2513$ in the $x-$ and $y-$ directions and $\delta_i\approx 1.005$ in the $z-$ direction. Notably, the ion inertial scales are greater than the dissipation scale (the spatial step sizes). As a result, reconnections because of both the Hall effect and the MPDATA-assisted residual dissipation are expected to be near-simultaneous in the presented simulations and onset with a steepening of current density. 

\begin{figure}[h]
\includegraphics[width=1\linewidth]{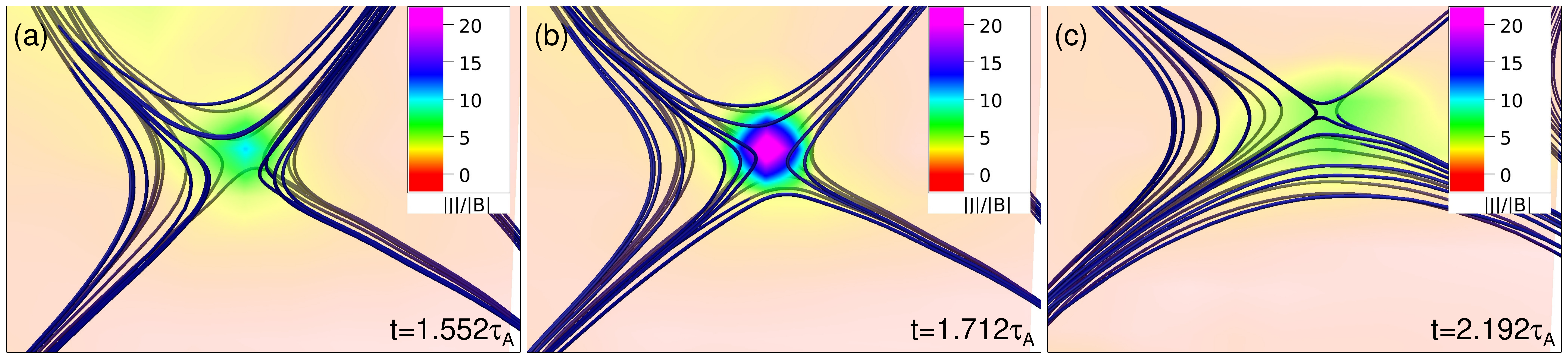}
\caption{\label{fig-cs} Variation of $|\textbf{J}|/|\textbf{B}|$ around the null point topology. This is the zoomed view of the same null point topology as shown in Figure \ref{fig1}(c) with blue MFL.}
\end{figure}

The simulation results presented herein correspond to a total run of 7000$\Delta t$  with the dimensionless temporal step size $\Delta t=16\times 10^{-4}$. The total simulated physical time is $7000\times\Delta t\times\tau_{A}=11.2\tau_A$. For convenience, hereafter (including figures) the time is presented in units of $\tau_A$.  
The $B_z$ and $v_z$ at the bottom boundary (at $z = 0$) are kept fixed throughout the simulations while all other field variables are allowed to vary. 
Noticeably, as the reconnection regions are not in the vicinity of the bottom boundary and are located at a certain height during the simulations, the scales remain completely resolved at the bottom boundary and plasma remains perfectly ideal at the boundary. Moreover, the z-component of velocity is set to be zero at the bottom boundary and the horizontal components of velocity develop as per the incompressibility condition. As a consequence, the employed boundary conditions mimic the line-tied effect at the bottom boundary \cite{jiang2021}. At all other boundaries, all variables including $B_z$, vary with time with their values at a given spatial location on the boundary being mapped from the immediate spatial neighbourhood. Importantly, the boundary condition used here is entirely different from the periodic boundary used in  ref. \cite{Sanjay2016} and allows for the generated flux rope to be anchored.

\section{Simulation results}
Our investigation toward the generation of a 3D MFR reveals repetitive 3D magnetic reconnections occurring at magnetic null points in both HMHD and MHD simulations. The presence and location of such null points are confirmed by utilizing the well-established and tested trilinear method of null detection in three-dimensional vector space; see  ref. \cite{Haynes} for details. Null detection shows that there are no null points present initially at $t=0$. The reconnections onset as the initial non-zero Lorentz force\footnote{Expression for the initial Lorentz force associated with the initial 3D bipolar sheared arcade is given in \ref{app}} (direction and magnitude around the PIL are shown in the panels (a) and (b) of Figure \ref{fig1}) pushes the oppositely directed segments of MFL toward each other to generate the neck (panels (c) and (d) of Figure \ref{fig1}). At the neck, the null points are detected and an instance of reconnection at the null points from each simulation (at $t=1.712\tau_A$) is presented in panels (c) and (d) of Figure \ref{fig1}. We also mark the spine of the nulls in Figure \ref{fig1} (c) and (d). 

To confirm the onset of reconnection at the null point, in Figure \ref{fig-cs}, we present the temporal variation of $|\textbf{J}|/|\textbf{B}|$ on a plane passing through the null point.  $|\textbf{J}|/|\textbf{B}|$ is generally taken as a proxy to locate regions of high current density, which are susceptible to reconnection \cite{Innoue21, Innoue23}. From panels (a)-(b), we notice that as the non-parallel field lines come in close vicinity, there is a significant enhancement in the value of the $|\textbf{J}|/|\textbf{B}|$ --- suggesting a sharp increase in the gradient of the magnetic field in the neighborhood of the neutral point. Consequently, the scales become under-resolved and magnetic reconnection initiates, causing a decrement in the value of  $|\textbf{J}|/|\textbf{B}|$ (panel (c)). It is noteworthy that in the absence of the neutral points in the initial magnetic field, the net topological degree \cite{wyper} is zero. Subsequently, careful analysis of the magnetic field vectors around null points reveal that the null point with blue MFLs has a topological degree -1 (MFLs approaching the null along the spine) while the null point with red MFLs has a topological degree +1 (MFLs receding from the null along the spine); see Figure \ref{fig1}(c) and (d). As a result, the net topological degree remains zero---suggesting that our simulations are in conformity with the principle of topological degree conservation \cite{wyper}. Determining the topological degree of all the nulls detected with the trilinear method is difficult and beyond the scope of this work, since the maximum number of null points detected in the HMHD case is $\approx 650$ and in the MHD case it is $\approx 500$.
 
Noticeably, the initial magnetic field configuration is symmetric about $x=\pi$ plane (Figure \ref{fig1}(a)), i.e., in the $x\in\{0,\pi\}$ and $x\in\{\pi,2\pi\}$ domain. Dynamical evolution and structures are the same about $x=\pi$; as evident from panels (c) and (d) of Figure \ref{fig1}. Therefore, for clear illustrations of the simulated dynamics, we choose to present the results of the simulations only in the $x\in\{\pi,2\pi\}$ domain in the subsequent figures.

\begin{figure}[h]
\includegraphics[width=1\linewidth]{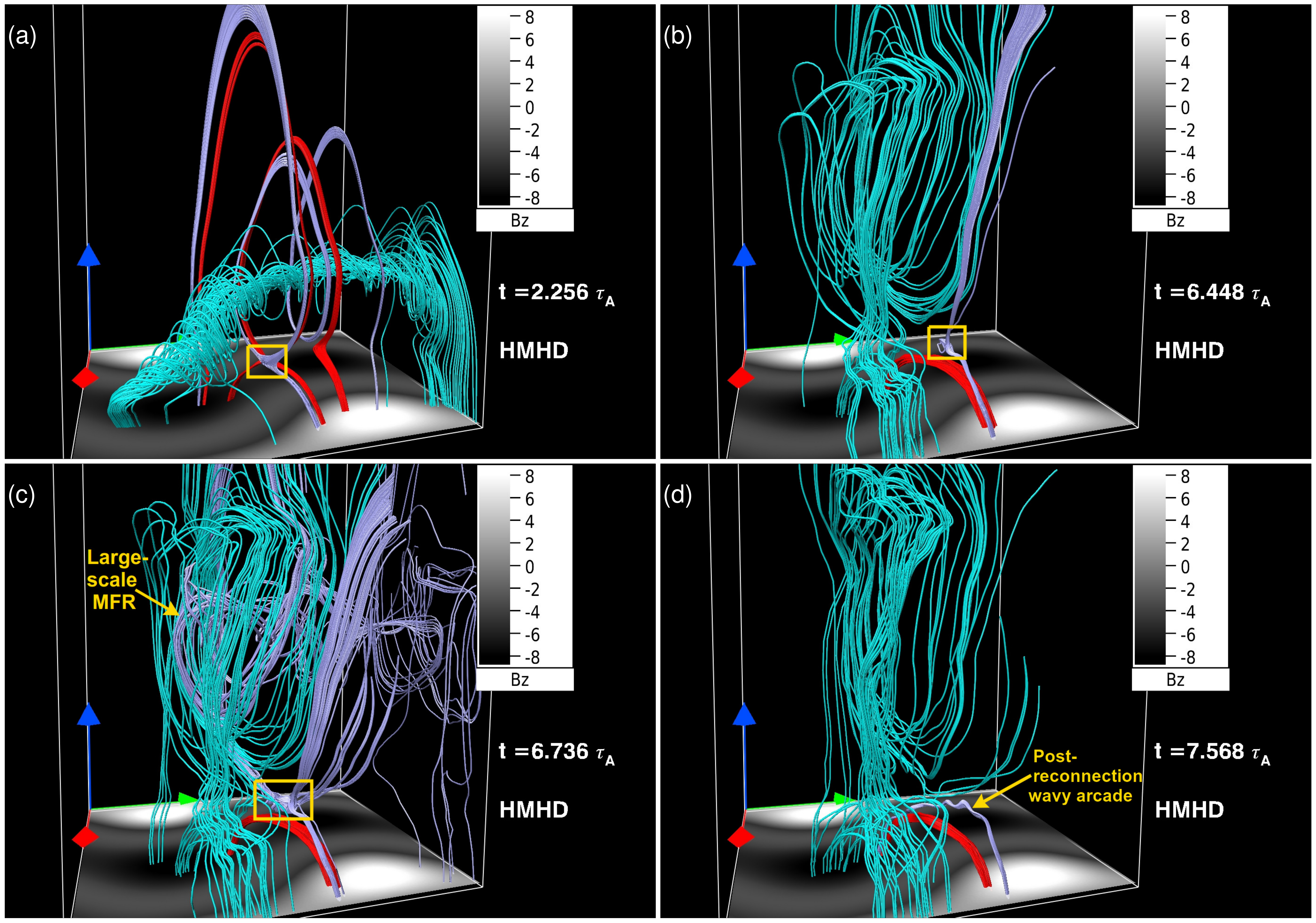}
\caption{\label{fig2}Snapshots from the HMHD evolution of a 3D MFR. (a) The two sets of field lines (red from the left and lavender from the right side) approaching each other (marked in a yellow rectangular box) below the MFR (cyan color) at $t=2.256 \tau_A$. (b) A structure formed by lavender color MFL (marked in the yellow rectangle) and the post reconnection arcade formed by red MFL at $t=6.448 \tau_A$. (c) Large-scale MFR formed by lavender color MFL and associated small-scale structure around reconnection site (marked in the yellow rectangle) at $t=6.736 \tau_A$. (d) Post-reconnection wavy arcade generated by lavender MFL at $t=7.568 \tau_A$.}
\end{figure} 

\begin{figure}[h]
\includegraphics[width=1\linewidth]{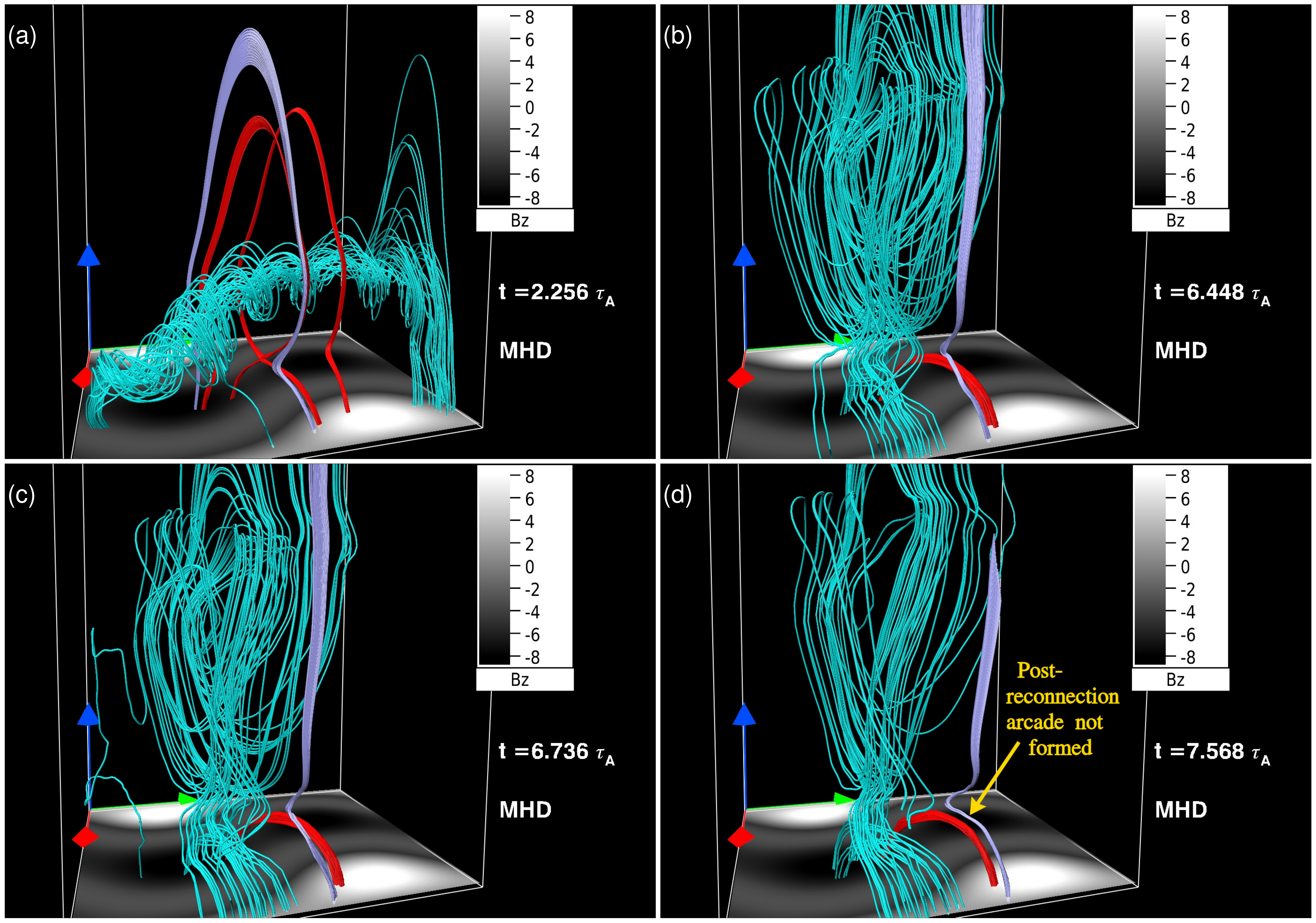}
\caption{\label{fig3}Snapshots from the MHD evolution of a 3D MFR. Panel (a) shows the MFR (cyan color) structure with two other sets of MFL (red and lavender) overlying the MFR. Notably, red and lavender color MFL are relatively farther compared to the HMHD instance of the same at $t=2.256 \tau_A$. Panels (b)-(d) highlight the morphology of lavender MFL. Noticeably, the lavender MFL does not exhibit any twisted structure formation near the reconnection site (cf. panels (b)-(d) of Figure \ref{fig2}.)}
\end{figure}

Figures \ref{fig2} and \ref{fig3} illustrate the instances of the evolution of selected MFL sets from the HMHD and MHD simulation respectively. 
To highlight the differences in the MFL dynamics around the reconnection sites and subsequent large-scale structural changes, we compare panels (a)-(d) of Figures \ref{fig2} and \ref{fig3}. Panels (a) of the respective figures depict a similar twisted MFR (cyan color), overlying stretched field lines, and the reconnection site below MFR during the HMHD and MHD simulations. In contrast to the similarity in the MFR, the red and lavender MFLs are seen to approach each other in HMHD (marked by the yellow rectangle in Figure \ref{fig2}(a)) at $t=2.256\tau_A$ but in MHD, they remain separated  and the red MFLs reconnect with themselves below the MFR (Figure \ref{fig3}(a)). Panels (b) of Figures \ref{fig2} and \ref{fig3} also show the similarity in the post-reconnection arcade constituted by the red MFLs. It is only the lavender MFLs that exhibit different structures during the HMHD and the MHD evolutions from $t=6.448\tau_A$ onwards, hence we focus mainly on the dynamics of the lavender MFLs further. In HMHD, the lavender MFLs form a large-scale MFR at $t=6.736\tau_A$ as shown by the yellow arrow in panel (c) of Figure \ref{fig2}. Notably, a region of twisted MFLs exists just below the MFR, as highlighted by the yellow rectangle (Figure \ref{fig2}(c)). Contrary to the HMHD, such formation of a large-scale MFR which has an associated small twisted structure below it is absent during the MHD evolution (see panels (c) of Figures \ref{fig2} and \ref{fig3}). In addition to these differences between the two simulations, we also notice the earlier development of post reconnection arcade during the HMHD evolution over MHD (see panels (d) of Figures \ref{fig2} and \ref{fig3}). Notably, the post reconnection arcade during the HMHD is wavy (at $t=7.568\tau_{A}$, Figure \ref{fig2}(d)) whereas no such structure was found during the MHD evolution. A faster development of a post-reconnection arcade clearly indicates faster reconnection dynamics in HMHD. Notably, the post-reconnection wavy arcade is twisted (Figure \ref{fig2}(d)). Interestingly, if the twisted magnetic field lines confine plasma, they are usually observed as a filament (against solar disk) or prominence (against solar limb) on the Sun. 

Notably, the extent of twisting in MFRs is generally associated with kink-instability {\cite{TK2005}} and therefore, we check the variation of averaged twist number with a time to explore this possibility. The twist number {\cite{liu+2016ApJ}} is defined as
\begin{equation}
T_{\mathit{w}} = \displaystyle\int_{L}\frac{(\nabla\times\bf{B}).\bf{B}}{4\pi B^{2}} dl,    
\end{equation}
where the integral follows the path of a magnetic field line from one footpoint to the other. Following {\cite{duan2022}}, we estimate the averaged twist number corresponding to the MFR volume with $|T_{\mathit{w}}|\geq 1$ and found the maximum values of averaged twist number to be 1.98 and 2 for the MHD and HMMD cases, respectively. Previous studies regarding the threshold value for kink-instability (see ref. \cite{duan2022} and the references therein) suggest that if the averaged twist number ranges between 1.5 to 2.5, the MFR is stable against kink. Therefore, the MFRs are expected to be kink-stable in our simulations.

 \begin{figure}[h]
  \centering
\includegraphics[width=1\linewidth]{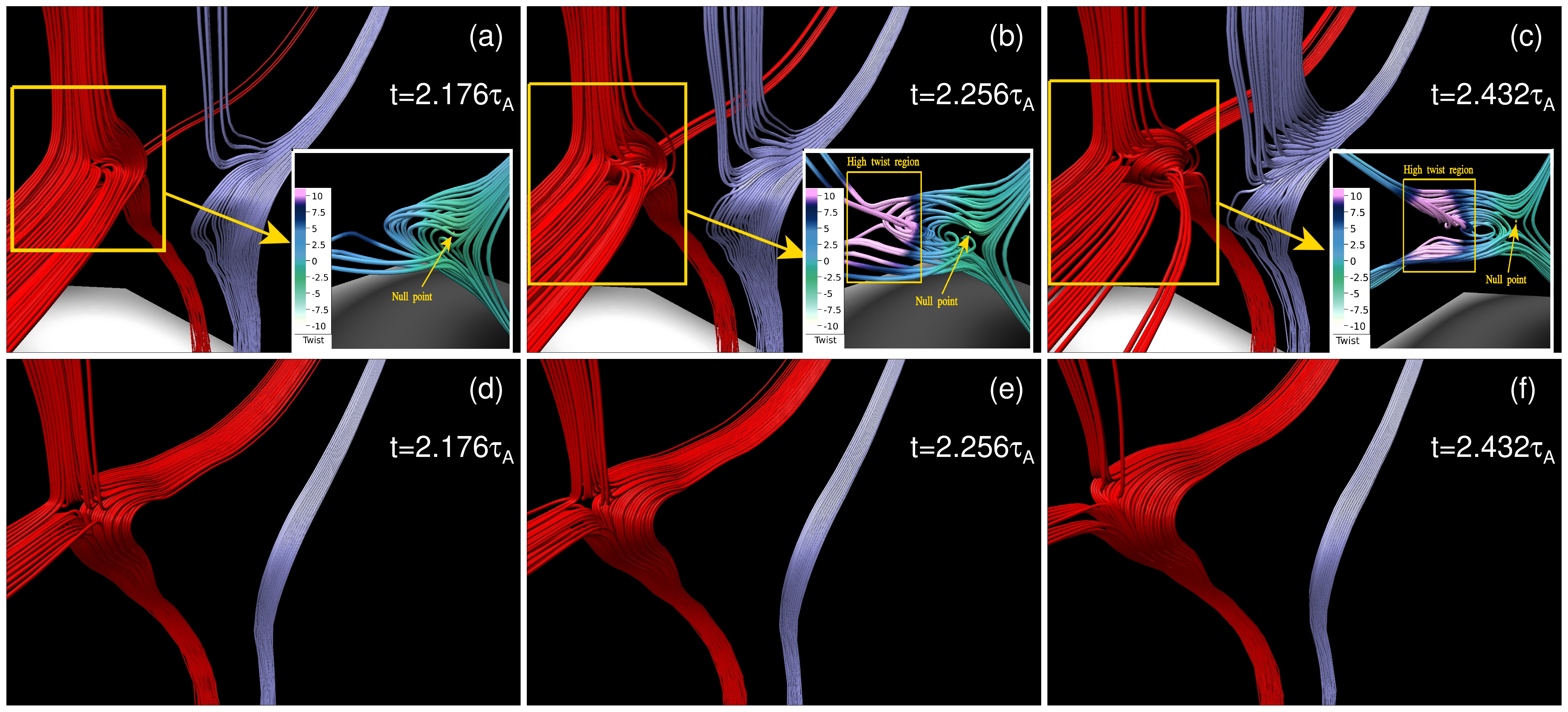}
  \caption{Detailed side view of the magnetic field lines marked in  Figure \ref{fig2} (a). Panels (a)-(c) depict the snapshots from the HMHD simulation and panels (d)-(f) depict the snapshots from the MHD simulation. 
Formation of twisted magnetic structure by red MFL is evident during the HMHD in panels (a)-(c), whereas no such structure is found during the MHD simulation(c.f. panels (d)-(f)). The inset images in panels (a)-(c) show the presence of a 3D null point and the development of a twist in the neighbourhood of the null point. MFL in inset images is the red MFL (marked in yellow rectangular box) color coded with the value of twist ($\alpha=\textbf{J}\cdot\textbf{B}/|\textbf{B}|^2$).}
  \label{fig4}
\end{figure}

\begin{figure}[h]
  \centering
\includegraphics[width=1\linewidth]{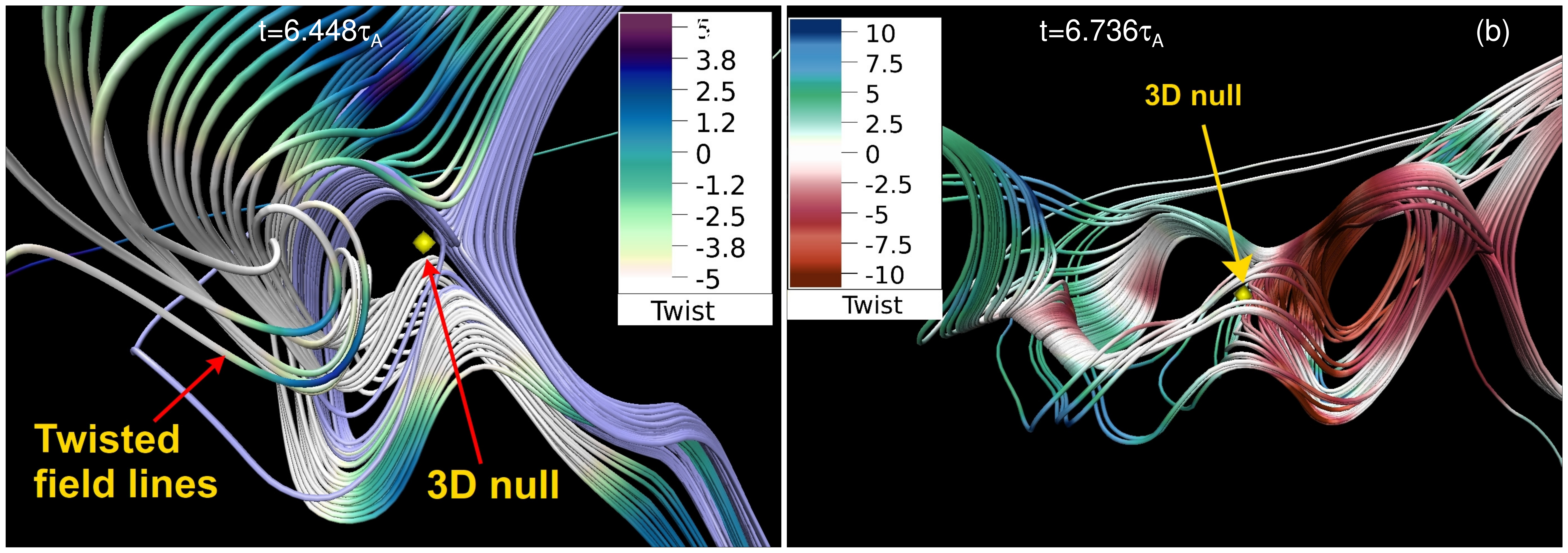}
  \caption{Panels (a) and (b) show the zoomed frontal view of the lavender MFL structure (marked in Figure \ref{fig2} (b) and \ref{fig2}(c)) highlighting the presence of a 3D null at $t=6.448 \tau_A$ and $t=6.448 \tau_A$ during the HMHD evolution. In panel (a) additional field lines (left to 3D null and color-coded with the twist value $\alpha$) show the twist value $\sim -5$ around the 3D null. Notably, in panel (b), the MFL has twist $\alpha\sim-10$ on right and $\alpha\sim(5-7.5)$ on left to the 3D null.}
  \label{fig5}
\end{figure}

In Figure \ref{fig4}, we present the temporal evolution of the zoomed-in view of the MFLs evolution in the region marked by the yellow rectangle in Figure \ref{fig2}(a). Snapshots from the HMHD and MHD simulations are depicted in panels (a)-(c) and (d)-(f) respectively. Inset images in panels (a)-(c) show the frontal view of red MFLs (color coded with the twist $\alpha=\textbf{J}\cdot\textbf{B}/|B|^2$) around the null point (yellow point marked by arrow). 
\begin{figure}[h]
  \centering
\includegraphics[width=0.75\linewidth]{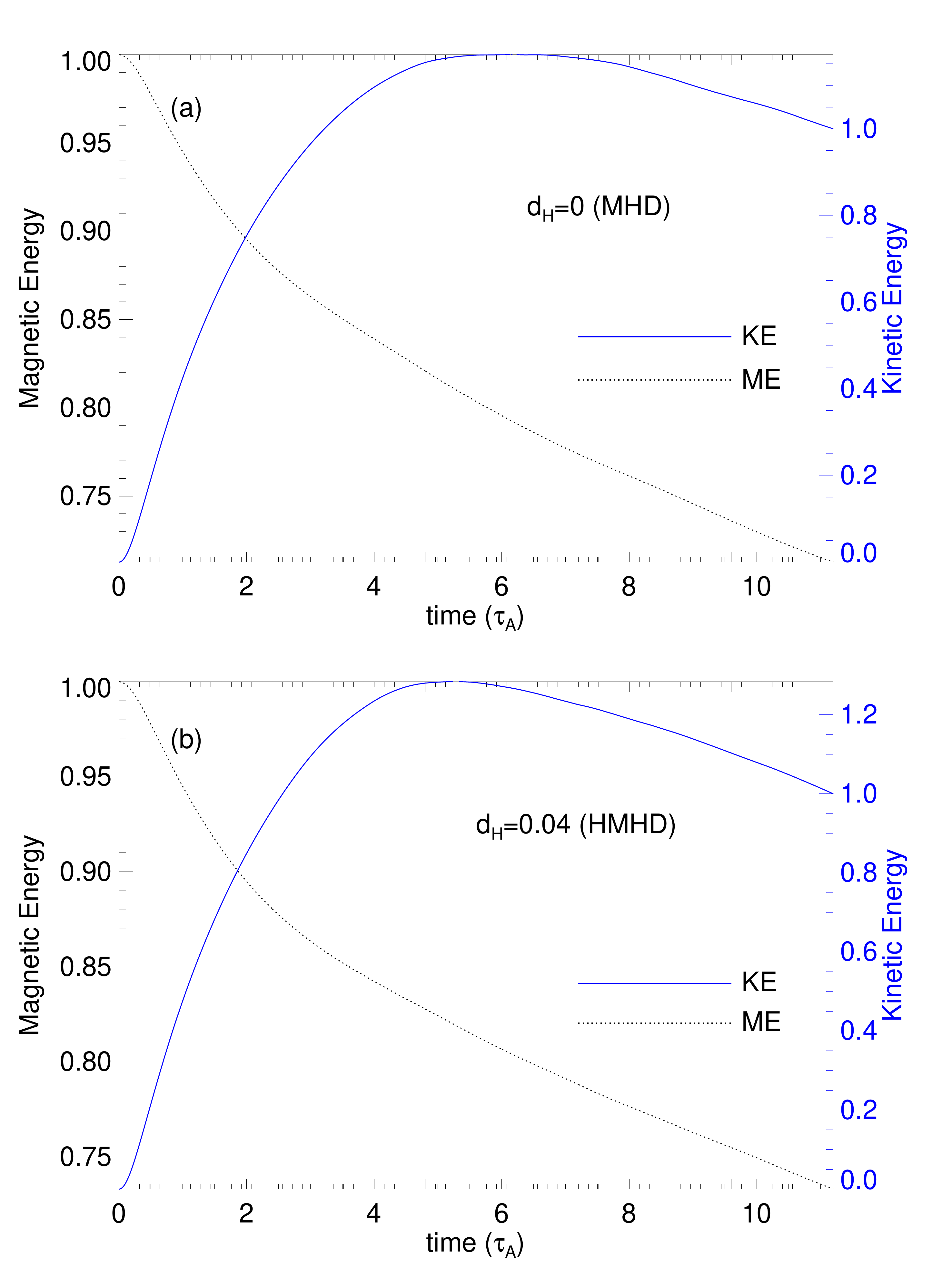}
  \caption{Panels (a) and (b) show the temporal variation of the volume-averaged magnetic (black dashed curve) and kinetic energies (blue solid curve) during the MHD and HMHD evolution respectively. }
  \label{fig6}
\end{figure}

Noticeably, during the HMHD simulation, the development of highly twisted MFL in the neighborhood of null point with the twist value $\alpha\approx 10$ is observed (the inset images in Figure \ref{fig4}(b) and (c)), but there are no such twisted MFL observed during the MHD. Furthermore, the lavender color MFLs form intermediate small-scale\footnote{Size compared to the large-scale MFR} structure during the HMHD evolution (marked within a yellow rectangle in Figure \ref{fig2}(b) and (c)). We present the zoomed-in frontal view of these structures in Figure \ref{fig5}. In panel (a), we have plotted additional field lines to the left of 3D null. These field lines are color coded with the value of twist and show the high twist around 3D null, i.e., $\alpha\approx -5$. This figure highlights the presence of null points in both the panels and a twisted mini flux rope in panel (b). The lavender color large-scale MFR (Figure \ref{fig2}(c)) is formed as a result of reconnections in the region shown in Figure \ref{fig5}(b). These small-scale, complex, and twisted structures surrounding the null points are associated with the formation of large-scale MFR during the HMHD but no such structures were found in the MHD case.
\par Temporal variation of the volume-averaged magnetic and kinetic energies is presented in panels (a) and (b) of Figure \ref{fig6}. The solid blue and dashed black curves represent the kinetic and magnetic energy variations respectively. Noticeably, the magnetic energy is decreasing identically in both simulations, hence the results are in agreement with the general theoretical expectation that the Hall effect does not cause changes in the magnetic energy dissipation rates \cite{Liu-hall}. Further, to rule out any dependency of the above results on resolution, auxiliary simulations (both MHD and HMHD) are performed on a grid $32\times32\times64$ while keeping all other parameters identical except $d_H$. The auxiliary simulations (animations provided as supplementary material) confirm a similar development of MFR along with the subsequent twisting of MFL in the near-neighborhood of the reconnection site (3D nulls) during the HMHD evolution.

\section{Summary and Discussion}

This paper presents a comparative study of flux rope evolution within the paradigm of magnetohydrodynamics and Hall magnetohydrodynamics numerical simulations. The rationale behind using two different frameworks of magnetized plasma modeling (with and without the Hall effect) comes from the possibility of arriving at a differentiator that could help to gain new insights into the dynamics of 3D MFRs. For this purpose, an initial 3D sheared bipolar magnetic arcade configuration is employed to initiate the numerical simulations. In both the MHD and HMHD cases, the initial plasma flow is set to zero with the line-tying condition on the bottom boundary, while other boundaries are treated open.

\par The initial non-zero Lorentz force drives the arcade configuration toward magnetic reconnection which subsequently generates 3D magnetic null points. These 3D nulls are precisely detected as locations that host reconnection to generate a 3D MFR in both the cases. Remarkably, for the reported nulls in this work, the net topological degree conservation is found. We remark that this result does not take into account all the magnetic nulls detected in the computational box. Such an analysis is presently beyond the scope of this work and importantly, does not directly relate to the main objective of this paper. We confirm the occurrence of reconnection at these 3D nulls essentially by analyzing the changes in magnetic field topology. Furthermore, the time evolution of magnetic field gradient at and around 3D nulls, characterized by the development of large values in $|\textbf{J}|/|\textbf{B}|$ followed by subsequent dissipation, corroborates well with our interpretation of reconnection. We also checked for the possibility of kink-instability in both MHD and HMHD cases. The averaged twist number corresponding to MFR volume with $|T_{\mathit{w}}|\geq 1$ turns out to be 2 and 1.98 for MHD and Hall MHD cases, which suggests that the generated 3D MFRs are kink stable.

\par A comprehensive comparison of the two simulations shows that the reconnections forming the MFR are identical, i.e., insensitive to the Hall forcing. However, their later evolution differs qualitatively as well as quantitatively. The earlier formation of post-reconnection arcades during the HMHD evolution implies faster dynamics over its MHD counterpart. Importantly, the Hall effect causes the twisting of magnetic field lines around the reconnection sites (3D nulls). Such twisting is unique to the HMHD evolution and affects the large-scale dynamics. Relevantly, ref. \cite{Pant_2018} have recently observed the plasma performing twisting/swirling motion while falling back on the Sun after a prominence eruption event. HMHD simulation results point toward the possibility that the helical/twisted field lines in the vicinity of reconnection sites may explain the observed twisting/swirling plasma motion during the prominence eruption. 

\par Both the MHD and HMHD simulations show an identical variation of the volume-averaged magnetic energy. This result is in accordance with the theoretical expectation that the Hall forcing being ideal should not change the magnetic energy rate \cite{Liu-hall}. In this paper, the MFR does not exist initially but gets generated through reconnections. The present simulations are constrained by the assumed incompressibility which restricts the thermodynamics of the MFRs. In future, we aim to advance the present simulations by relaxing the incompressibility condition to make the simulations more realistic.

\bigskip
\bigskip

\noindent \textbf{SUPPLEMENTARY MATERIAL}\\
See supplementary material for the comparison of the HMHD and MHD evolution of the 3D MFR.

\bigskip
\bigskip

\noindent \textbf{ACKNOWLEDGEMENT}\\
We thank the anonymous referees for providing valuable inputs and suggestions to increase the scientific content and readability of this paper. The simulations are performed using the 100TF cluster Vikram-100 at Physical Research Laboratory, India. We wish to acknowledge the visualization software VAPOR (\url{www.vapor.ucar.edu}), for generating relevant graphics in this paper. For the null detection, we have used a python code based on the trilinear method, developed by Federica Chiti, David Pontin, Roger Scott and available at \url{https://zenodo.org/record/4308622#.YByPRS2w0wc}.

\bigskip
\bigskip

\noindent \textbf{CONFLICT OF INTEREST STATEMENT} \\
The authors have no conflicts to disclose.

\bigskip
\bigskip
  
\noindent \textbf{DATA AVAILABILITY}\\
 The simulation data that supports the finding of this study are available from the corresponding author upon reasonable request.
\bigskip
\bigskip

\appendix

\section{Lorentz force}
\label{app}
Initial Lorentz force corresponding to the initial 3D bipolar sheared field is given as following:

\begin{eqnarray}
(\textbf{J}^*\times\textbf{B}^*)_x&=&s_0\left(\frac{1}{2s_0}-s_0\right)\sin x~\cos x~ \sin^2 y~ \exp\left(-\frac{2z}{s_0}\right)-\frac{1}{2}\sin^2 x~\sin^2 y~\exp\left(-\frac{2z}{s_0}\right)\nonumber\\
&&+\left(s_0\left({a_x}^2-\frac{{a_z}^2}{s_0}\right)\sin y~+\frac{\sqrt{{a_x}^2-{a_z}^2}}{2}(\cos y-\sin y)\right)\sin x~\sin (a_x x)\times\nonumber\\
&&\exp\left(-\frac{(1+a_z)z}{s_0}\right)\nonumber\\
&&+\left(s_0 a_x\left(\frac{1}{2s_0}-s_0\right)\cos x - \frac{a_x}{2}\sin x\right)\cos (a_x x)~ \sin y~\exp\left(-\frac{(1+a_z)z}{s_0}\right)\nonumber\\
&&+\left(\frac{s_0 a_x}{2}\left(s_0 {a_x}^2-\frac{{a_z}^2}{s_0}\right)-\frac{a_x({a_x}^2-{a_z}^2)}{2}\right)
~\sin (2a_x x)~\exp\left(-\frac{2a_z z}{s_0}\right)\nonumber\\
&&+\frac{1}{4}\sin x ~(\sin y -\cos y)(\cos x~ \sin y - \sin x ~\cos y)~\exp\left(-\frac{2z}{s_0}\right)\nonumber\\
&&+\frac{a_x\sqrt{{a_x}^2-{a_z}^2}}{2}\cos (a_x x)(\cos x~ \sin y - \sin x ~\cos y)~\exp\left(-\frac{(1+a_z)z}{s_0}\right)\nonumber\\
(\textbf{J}^*\times\textbf{B}^*)_y&=&s_0\left(\frac{1}{2}-s_0\right)\sin^2 x~\sin y~ \cos y~ \exp\left(-\frac{2z}{s_0}\right)+\sin x~\cos x~\sin^2 y~\exp\left(-\frac{2z}{s_0}\right)\nonumber\\
&& +\frac{1}{4}\sin x~(\sin y - \cos y)(\sin x ~\sin y - \cos x ~\sin y)\exp\left(-\frac{2z}{s_0}\right)\nonumber\\
&&+\frac{a_x\sqrt{{a_x}^2-{a_z}^2}}{2}\cos (a_x x)(\sin x~ \sin y - \cos x~ \sin y)\exp\left(-\frac{(1+a_z)z}{s_0}\right)\nonumber\\
&&+\left(\frac{a_z}{2}(\sin y - \cos y)-a_z\sqrt{{a_x}^2-{a_z}^2}\right) \sin x \sin (a_x x)~\exp\left(-\frac{(1+a_z)z}{s_0}\right)\nonumber\\
&&+\left(a_z\left(\frac{1}{2}-s_0\right)\sin x~\cos y + a_x \cos x~\sin y\right) \cos (a_x x)~\exp\left(-\frac{(1+a_z)z}{s_0}\right)\nonumber\\
(\textbf{J}^*\times\textbf{B}^*)_z&&=\left(\frac{\sin x}{4s_0}-\left(\frac{1}{2s_0}-s_0\right)\frac{\cos x}{2}\right)\sin y~(\sin x~\sin y-\cos x~\sin y)~\exp\left(-\frac{2z}{s_0}\right)\nonumber\\
&&+\left(\frac{1}{2}\left(\frac{1}{2}-s_0\right)\sin x~\cos y+\frac{1}{2s_0}\cos x~\sin y\right)(\cos x ~\sin y-\sin x~ \cos y)\times\nonumber\\ 
&&\exp\left(-\frac{2z}{s_0}\right)\nonumber\\
&&+\frac{a_x\sqrt{{a_x}^2-{a_z}^2}}{2s_0}\sin (a_x x)(\cos x ~\sin y-\sin x~ \cos y)~\exp\left(-\frac{(1+a_z)z}{s_0}\right)\nonumber\\
&&-\left(\sqrt{{a_x}^2-{a_z}^2}\left(\frac{1}{2}-s_0\right)\sin x ~\cos y+\frac{\sqrt{{a_x}^2-{a_z}^2}}{s_0}\cos x~\sin y\right)\sin (a_x x)\times\nonumber\\
&&\exp\left(-\frac{(1+a_z)z}{s_0}\right)\nonumber\\
&&+\left(\frac{a_z({a_x}^2-{a_z}^2)}{s_0}+a_z\left(\frac{{a_z}^2}{s_0}-s_0~{a_x}^2\right)\right)\sin^2 (a_x x)~\exp\left(-\frac{2a_z z}{s_0}\right)\nonumber\\
&&+\frac{1}{2}\left(\frac{{a_z}^2}{s_0}-s_0 {a_x}^2\right)\sin (a_x x) (\sin x~ \sin y-\cos x~\sin y)~\exp\left(-\frac{(1+a_z)z}{s_0}\right)\nonumber\\
&&+\left(\frac{a_z}{2s_0}\sin x-a_z\left(\frac{1}{2s_0}-s_0\right)\cos x\right)\sin (a_x x)~\sin y~\exp\left(-\frac{(1+a_z)z}{s_0}\right)\nonumber
\end{eqnarray}

\section*{References} 
\bibliography{ref.bib}

\end{document}